\documentclass[aps,prx,reprint,twocolumn,floatfix,superscriptaddress,amsmath,amssymb,amsfonts,nofootinbib,balancelastpage]{revtex4-2}

\usepackage{url}
\usepackage{bm}
\usepackage{graphicx}
\usepackage{amsmath}
\usepackage{amstext}
\usepackage{amssymb}
\usepackage{amsfonts}
\usepackage{amsbsy}
\usepackage{verbatim}
\usepackage{color}
\usepackage{floatrow}
\usepackage{float}
\usepackage{tikz}
\usepackage{gensymb}
\usepackage{textcomp}
\usepackage{enumitem}
\usepackage[version=3]{mhchem}
\usepackage{afterpage}
\usepackage{pbox}
\usepackage{dsfont}
\usepackage{siunitx}
\usepackage{stackengine,wasysym,scalerel}
\usepackage{latexsym}
\usepackage{cancel}
\usepackage{comment}
\usepackage{array, multirow, bigdelim, makecell, booktabs}
\usepackage[misc]{ifsym}
\usepackage{sidecap}
\usepackage{mathrsfs}

\usepackage{color}
\usepackage{pifont}

\usepackage[normalem]{ulem}
\newcommand{\pdagger}{\phantom{\dagger}}
\usepackage{siunitx}
\usepackage{physics}
\usepackage{soul}
\usepackage[colorlinks=true,
            urlcolor=blue,
            citecolor=blue,
            linkcolor=blue]{hyperref}
            
\usepackage{pdfpages}
\usepackage{pgffor}
\makeatletter
\AtBeginDocument{\let\LS@rot\@undefined}
\makeatother

\usepackage{xcolor}

\begin{document}

\title{Chiral and bond-ordered phases in a triangular-ladder superconducting-qubit quantum simulator}

\author{Matthew Molinelli}\thanks{These authors contributed equally to this work.}
\affiliation{Department of Electrical and Computer Engineering, Princeton University, Princeton, NJ 08544, USA}
\author{Joshua C. Wang}\thanks{These authors contributed equally to this work.}
\affiliation{Department of Electrical and Computer Engineering, Princeton University, Princeton, NJ 08544, USA}
\author{Jeronimo G. C. Martinez}
\affiliation{Department of Electrical and Computer Engineering, Princeton University, Princeton, NJ 08544, USA}
\author{Sonny Lowe}
\affiliation{Department of Electrical and Computer Engineering, Princeton University, Princeton, NJ 08544, USA}
\author{Andrew Osborne}
\affiliation{Department of Electrical and Computer Engineering, Princeton University, Princeton, NJ 08544, USA}
\author{Rhine Samajdar}
\email{rhine\_samajdar@princeton.edu}
\affiliation{Department of Electrical and Computer Engineering, Princeton University, Princeton, NJ 08544, USA}
\author{Andrew A. Houck}
\email{aahouck@princeton.edu}
\affiliation{Department of Electrical and Computer Engineering, Princeton University, Princeton, NJ 08544, USA}

\begin{abstract}
Many-body systems with strong interactions often exhibit macroscopic behavior markedly absent in single-particle or noninteracting limits. Such emergent phenomena are well exemplified in lattice Hubbard models, where the interplay between interactions, geometric frustration, and magnetic flux gives rise to rich physics.
Superconducting qubits naturally enable analog quantum simulation of Bose-Hubbard models, while offering tunable parameters, site-resolved control, and rapid experimental repetition rates.
Here, we study a superconducting-qubit device that realizes the Bose-Hubbard model on a triangular-ladder lattice. By tuning the magnitude and sign of couplings, we engineer a synthetic magnetic flux to characterize the resulting half-filling ground state for various parameter regimes.
We measure observables analogous to current-current correlators and bond kinetic energies, finding signatures consistent with chiral superfluids, Meissner superfluids, and bond-ordered insulators.
Our results establish superconducting circuits as a platform for robustly probing quantum phases of matter in frustrated Bose-Hubbard systems, even in strongly correlated and gapless regimes.
\end{abstract}

\maketitle

\section*{Introduction}

Strongly interacting many-body systems display a wealth of emergent phenomena that cannot be inferred from the properties of their individual constituents alone~\cite{anderson1972more}. 
As an example, consider a system of particles on a periodic lattice. In the absence of interactions, the physics is easily understood: Bloch's theorem can be used to calculate band structures and eigenstates, and many properties follow directly from this noninteracting low-energy spectrum~\cite{kittel2004introduction}. 
In contrast, strongly interacting lattice systems exhibit rich emergent behavior wherein interactions result in nontrivial many-body scattering processes and collective modes, often stabilizing macroscopic phases that have no single-particle analog~\cite{sachdev2023quantum}. 
Since the intrinsic quantum correlations render these regimes exponentially difficult to simulate classically~\cite{lloyd1996universal}, quantum simulation~\cite{feynman1982simulating} offers valuable opportunities for progress.

The Hubbard model~\cite{hubbard1964electron}, a workhorse of modern condensed matter physics, provides a useful paradigm to probe a variety of such emergent phenomena in strongly correlated materials~\cite{arovas2022hubbard}.
In bosonic systems, described by the Bose-Hubbard model, the delicate competition between the lattice tunneling $J$ and the onsite interaction $U$ yields a phase diagram that depends sensitively on not only $J/U$ but also the filling $\rho$~\cite{Sachdev2011}. At very low filling, one simply recovers a perturbative connection to single-particle behavior. On the other hand, at integer filling, strong interactions lead to Mott-insulating states while weak interactions favor superfluid order~\cite{capello2007superfluid}. 
Between these limits---at intermediate fillings---additional control parameters, such as geometric frustration and magnetic fluxes, can drive qualitatively distinct behavior, especially in reduced dimensionality where the effects of interactions are maximized~\cite{giamarchi2003quantum}.

\begin{figure*}[t]
    \includegraphics[width=\textwidth]{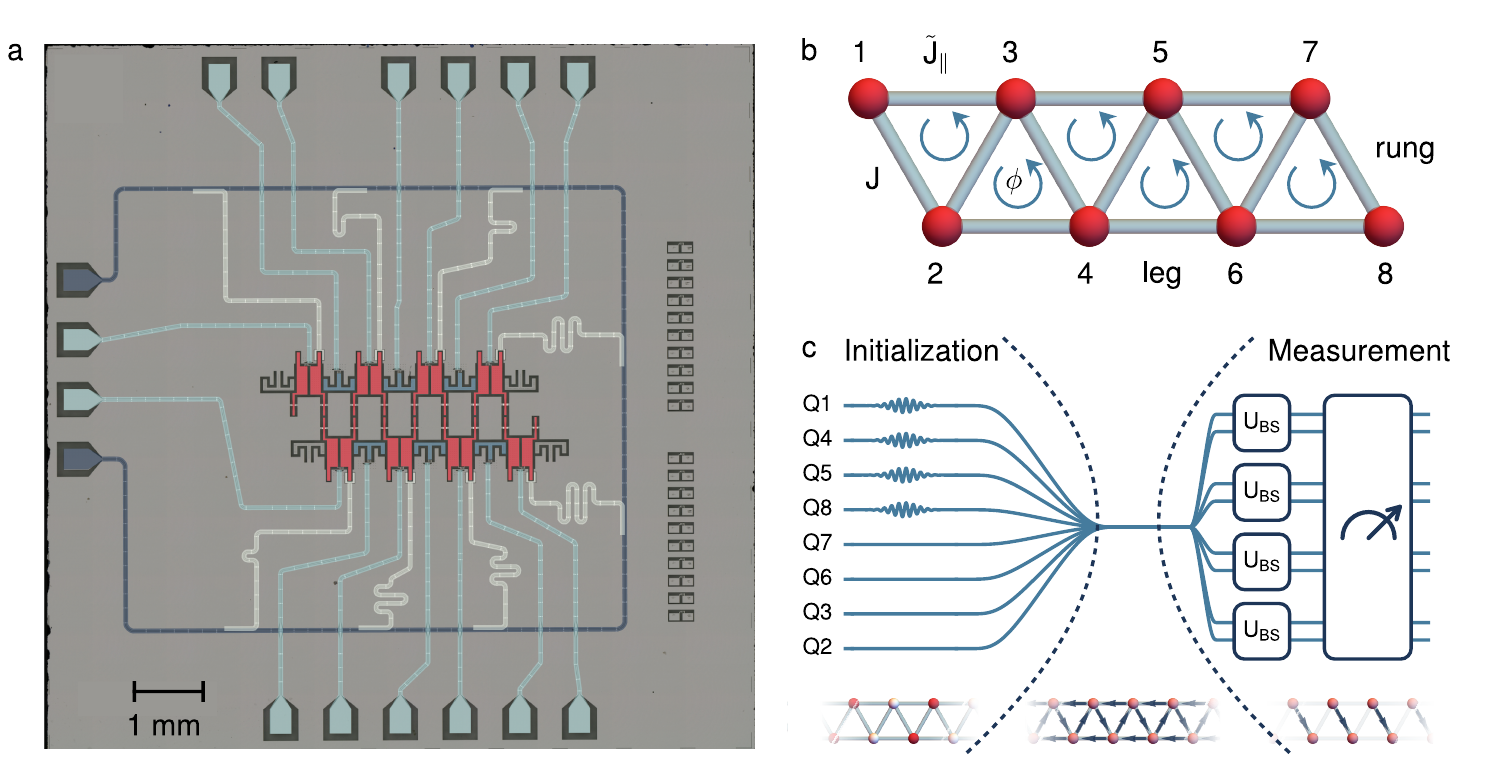}

    \caption{Realization of the triangular Bose-Hubbard ladder with transmon qubits. \textbf{(a)} False-color image of the 8-qubit triangular ladder with transmon qubits shaded in red and transmon couplers shaded in blue. Readout resonators and flux-bias lines are shaded in white and light blue, respectively. \textbf{(b)} Schematic of the 8-qubit triangular ladder. $J$ and $\tilde{J}_{\parallel}$ denote hoppings along the diagonals (rungs) and parallel edges (legs), respectively. Each plaquette is additionally threaded with a synthetic magnetic flux $\phi$. \textbf{(c)} Full experimental sequence. The four qubits with the highest energy are excited before the adiabatic ramp onto resonance. Arbitrary pairs of spatially adjacent qubits are isolated to characterize the state in two-qubit subspaces. Lattice schematics (bottom) depict the initialized state at half filling, the superfluid state after the ramp, and the isolated state during the measurement protocol. }
    \label{fig:fig1}
\end{figure*}

A simple geometry in which these effects of frustration and magnetic flux can be studied is the triangular-ladder lattice, composed of two coupled one-dimensional (1D) chains forming elementary plaquettes (Fig.~\ref{fig:fig1}b).
In the Bose-Hubbard model on such a ladder, competing hopping processes on the triangular geometry give rise to kinetic frustration~\cite{haerter2005kinetic, morera2022hightemperature,samajdar2023nagaoka,sb2}, resulting in phases beyond conventional superfluids and Mott insulators~\cite{Xu2023FrustrationDopingMagnetism,Prichard2024SpinPolarons,Li2023FrustratedChiralDynamics}.
Additionally, applying a synthetic magnetic field can strongly modify the lattice properties, leading to phenomena such as flat band dispersions~\cite{Martinez_2023,rosen2025flatbanddelocalizationemulatedsuperconducting}, flux-dependent chiral transport~\cite{Roushan2017Chiral}, and the emergence of interaction-driven phases~\cite{Impertro2025BosonicFluxLadder}.

Recent advances in analog quantum simulation have made it possible to implement and probe these Hamiltonians in the laboratory~\cite{altman2021quantum}.  Superconducting transmon~\cite{transmon} platforms are well poised for analog quantum simulation due to their native implementation of the Bose-Hubbard model as well as the flexibility of Hamiltonian engineering. Transmon devices admit tunable couplings~\cite{YanTunableCoupling}, strong interactions, and rapid experimental repetition rates, enabling the realization and characterization of interacting lattice models in regimes that are difficult to access classically. In the limit of weak anharmonicity and near-resonant coupling, a system of coupled transmons is naturally described by an attractive Bose-Hubbard Hamiltonian~\cite{Ma2019DissipativelyStabilizedMott}, where photons play the role of interacting bosons hopping between lattice sites. This mapping has enabled much experimental progress in exploring strongly correlated bosonic systems using superconducting circuits~\cite{Wang2024FractionalQuantumHall,Roushan2017Spectroscopic,Andersen2025ThermalizationCriticality}.

In this work, we realize an experimental implementation of the triangular-ladder Bose-Hubbard Hamiltonian with a synthetic gauge field on a superconducting-qubit device. By engineering both the magnitudes and signs of nearest-neighbor couplings, we implement a tunable flux of either $0$ or $\pi$ per plaquette and probe the ground-state properties within fixed particle-number sectors. We observe a variety of exotic many-body phases, including states with currents along the legs of the ladder (Meissner superfluids), states that spontaneously break a $\mathbb{Z}_2$ parity symmetry (chiral superfluids), and states that break translational symmetry (bond-ordered phases)~\cite{giamarchi}.
Our platform provides site-resolved access to observables analogous to current-current correlators and bond operators \cite{site_resolved_currents,MarkDWave}, allowing direct measurement of order parameters that distinguish these regimes. By mapping the ground-state response as a function of interaction strength, flux, and filling, and comparing our measurements to theoretical predictions, we demonstrate robust signatures of flux-induced chiral order and bond modulation in this frustrated geometry. These results thus showcase superconducting lattices as a flexible and controllable testbed for exploring the interplay of synthetic flux and geometric frustration in strongly interacting bosonic systems.

\section*{Device and Model}

The sites of our lattice are $N=8$ flux-tunable transmon qubits arranged in a quasi-1D triangular-ladder geometry, as shown in Fig.~\ref{fig:fig1}a, with the qubits highlighted in red in the false-color image. Adjacent qubits are capacitively coupled along the diagonals of the ladder, which we refer to as ``rungs'', with an average hopping of $J/2\pi = 6.1\ \mathrm{MHz}$. Along the top and bottom edges of the ladder, which we call ``legs'', qubits are coupled via a second set of flux-tunable transmons that act as couplers (shown in blue). Tuning the coupler frequency mediates hopping between the qubits in the range $J_{\parallel}/2\pi = \SI{2.5}{\MHz}$ to $\SI{20.4}{\MHz}$. When the coupler frequency is tuned below the qubit frequency, the effective hopping becomes negative, accessing values $J_{\parallel}/2\pi = \SI{-17.3}{\MHz}$ to $\SI{-7.0}{\MHz}$.

Our qubits are tunable from $\omega_q/2\pi = \SI{3.5}{\GHz}$ to $\SI{4.4}{\GHz}$ and are each coupled to a readout resonator with frequencies between $\omega_r/2\pi = \SI{7.1}{\GHz}$ and $\SI{7.6}{\GHz}$. In this work, we initialize and measure the qubits at distinct frequencies, but operate the lattice at a resonance frequency of either $\SI{3.85}{\GHz}$ or $\SI{4.30}{\GHz}$. The transmon qubits in our system are well approximated as weakly anharmonic bosonic oscillators~\cite{transmon}. The couplers are likewise transmons that mediate excitation transfer between neighboring qubits through a second-order hopping process. The device is therefore described by an interacting Bose-Hubbard model with the Hamiltonian

\begin{equation}
\label{eqn:hamiltonian}
\begin{aligned}
\frac{H}{\hbar} = &\sum_{j=1}^N \left(\omega^{\pdagger}_j n^{\pdagger}_j + \frac{U_j}{2} n^{\pdagger}_j(n^{\pdagger}_j - 1) \right) \\ 
- &\sum_{j=1}^{N-1} J_{j} \left(a_j^\dagger a_{j+1}^{\pdagger} + a_{j+1}^\dagger a_j^{\pdagger}\right)\\
+ &\sum_{j=1}^{N-2} \tilde{J}_{\parallel,j} \left(a_j^\dagger a_{j+2}^{\pdagger} e^{\mathrm{i} \phi} + a_{j+2}^\dagger a_j ^{\pdagger}e^{-\mathrm{i} \phi}\right).
\end{aligned}
\end{equation}
Here, $a_j$ represents the bosonic annihilation operator on lattice site $j$, and $n^{\pdagger}_j \equiv a_j^\dagger a^{\pdagger}_j$ is the corresponding number operator. $U_j$ denotes the onsite interaction arising from the transmon anharmonicity, with average value $U/2\pi = -186.1\ \mathrm{MHz}$. $J_j > 0$ is the rung hopping between sites $j$ and $j+1$, while $\tilde{J}_{\parallel,j}>0$ is the leg hopping between sites $j$ and $j+2$, which relates to the physical coupling strength as $J_{\parallel,j} = \tilde{J}_{\parallel,j} \mathrm{e}^{\mathrm{i}\phi}$.

Each triangular plaquette (see Fig.~\ref{fig:fig1}b) comprises two rungs with hopping $J$ and one leg with hopping $\tilde{J}_{\parallel}$, so the phase of the synthetic magnetic field through the loop is controlled by the sign of $J_{\parallel}$. Hence, we define the flux as $0$ for $J_{\parallel} > 0$ and as $\pi$ for $J_{\parallel} < 0$. Note that we have chosen a gauge in which the Peierls phase $\phi$ is allocated entirely to the leg hopping. Experimentally, we set $\phi = 0$ or $\pi$ by changing the sign of $J_{\parallel,j}$ as described above.

The Hamiltonian~\eqref{eqn:hamiltonian} commutes with the total-particle-number operator,
$[H, \sum_j n_j]=0,$ and therefore possesses a global U$(1)$ symmetry.  As a consequence, one may choose eigenstates to have definite particle number $\mathcal{N}$.
For repulsive onsite interactions ($U$\,$>$\,$0$), the competition between kinetic energy  and interactions yields the familiar Mott-insulator versus superfluid phenomenology: at commensurate (integer) filling and sufficiently large $U/J$, the ground state is a gapped Mott insulator with suppressed number fluctuations, while for $J \gg U$ the kinetic term dominates and the system develops long-range phase coherence (i.e., superfluid order)~\cite{Sachdev2011}. This is diagnosed by off-diagonal long-range order (ODLRO) of the one-body density matrix, which corresponds to a nonvanishing asymptotic value of $\langle  a_i^\dagger  a_j^{\pdagger} \rangle$ as $|i-j|\to\infty$.

The attractive case ($U$\,$<$\,$0$) is qualitatively different.  A strong onsite attraction minimizes energy by bunching all particles on a single site, so the natural Fock states at large $|U|$ are localized states, $|\mathcal{N}_j\rangle\propto(\hat a_j^\dagger)^\mathcal{N}|0\rangle$. Crucially, the translationally invariant  ground state, which is the symmetric superposition in this fixed-number sector, has vanishing off-diagonal one-body correlations between distinct sites,
$\langle  a_i^\dagger  a_j^{\pdagger} \rangle$\,$=$\,$0$ for $i$\,$\neq$\,$j$, and therefore, no ODLRO.  In other words, 
the exact ground state exhibits no global phase coherence and is not a superfluid.

In order to probe superfluid physics with our system, we instead consider the ground state of the negative Hamiltonian. Negating the Hamiltonian implements the transformation
\begin{equation}
\label{eqn:sign_trick}
    -H(\phi,U)=H(\phi+\pi,-U).
\end{equation}
The plaquette phase is shifted by $\pi$ because all couplings are negated and each plaquette contains an odd number of edges. To access the ground state of the \textit{repulsive} Hamiltonian at $\pi$ flux, we first prepare the highest excited state of the \textit{attractive} Hamiltonian at $0$ flux and then exploit the sign-flip mapping of Eq.~\eqref{eqn:sign_trick}. Since we can control the sign of $J_{\parallel}$, this procedure allows a negative-temperature state of the experimentally realized attractive Bose-Hubbard model to faithfully represent the low-energy physics of the repulsive model~\cite{Chen_2023}, a strategy we employ throughout this work.

\section*{Results}

Our experiments focus on the half-filling sector corresponding to four excitations in an eight-site ladder. At this filling, the lattice is neither dilute nor fully occupied, which ensures that kinetic and interaction energies compete most strongly.

The experimental procedure consists of two stages: state preparation and measurement, shown in Fig.~\ref{fig:fig1}c. State preparation involves exciting four out of the eight qubits and ramping all qubits onto resonance to prepare the half-filling ground state of the repulsive Hamiltonian. Next, we let the system evolve under a specific Hamiltonian, which implements a calibrated unitary operation, to rotate the desired observables to the computational basis before performing simultaneous projective readout on all eight qubits to collect bitstring outcomes. Repeated runs then yield the expectation values and correlation functions used to construct the order parameters reported below.

In order to probe ODLRO, we characterize the prepared eigenstate by measuring pairwise particle-number current operators along rungs of the lattice. 
For a rung connecting sites $j$ and $j+1$, the Hermitian current operator is given by
\begin{equation}
\mathcal{J}^{\pdagger}_{j}= \mathrm{i}J^{\pdagger}_{j}\left(a_j^\dagger a_{j+1}^{\pdagger} - a_{j+1}^\dagger a_j^{\pdagger}\right),
\end{equation}
so that its expectation value is directly related to the nearest-neighbor coherences, $
\langle\mathcal{J}^{\pdagger}_j\rangle = 2J^{\pdagger}_j\,\mathrm{Im}\langle a_j^\dagger a^{\pdagger}_{j+1}\rangle.
$
A nonzero $\langle\mathcal{J}_j\rangle$ therefore signals finite short-range phase coherence (a necessary condition for ODLRO), whereas insulating states with suppressed off-diagonal correlations have $\langle\mathcal{J}_j\rangle\approx0$.

As the current operator generates population exchange between neighboring sites, its expectation value can be inferred from the population imbalance after controlled time evolution. To that end, we first time-evolve each pair of qubits individually under the Hamiltonian $H^{\pdagger}_{\text{BS},j}$\,$=$\,$\hbar J^{\pdagger}_{j} (a_j^\dagger a_{j+1}^{\pdagger}$\,$+$\,$a_{j+1}^\dagger a_j^{\pdagger})$ for a ``beamsplitter'' time $t_{\text{BS}} = \pi/4J$ and estimate the current $\mathcal{J}_j$ at time $t = 0$ by~\cite{site_resolved_currents}
\begin{equation}
\label{eqn:current}
    \mathcal{J}^{\pdagger}_j(0) \approx J^{\pdagger}_j\big [n^{\pdagger}_j(t^{\pdagger}_{\text{BS}}) - n^{\pdagger}_{j+1}(t^{\pdagger}_{\text{BS}})\big].
\end{equation}
This relation becomes exact in the hard-core boson limit, where the system reduces to the frustrated spin-1/2 quantum XX model~\cite{sato2011competing,mishra2013quantum, mishra2014supersolid}.

\begin{figure}[t]
    \includegraphics[width=\linewidth]{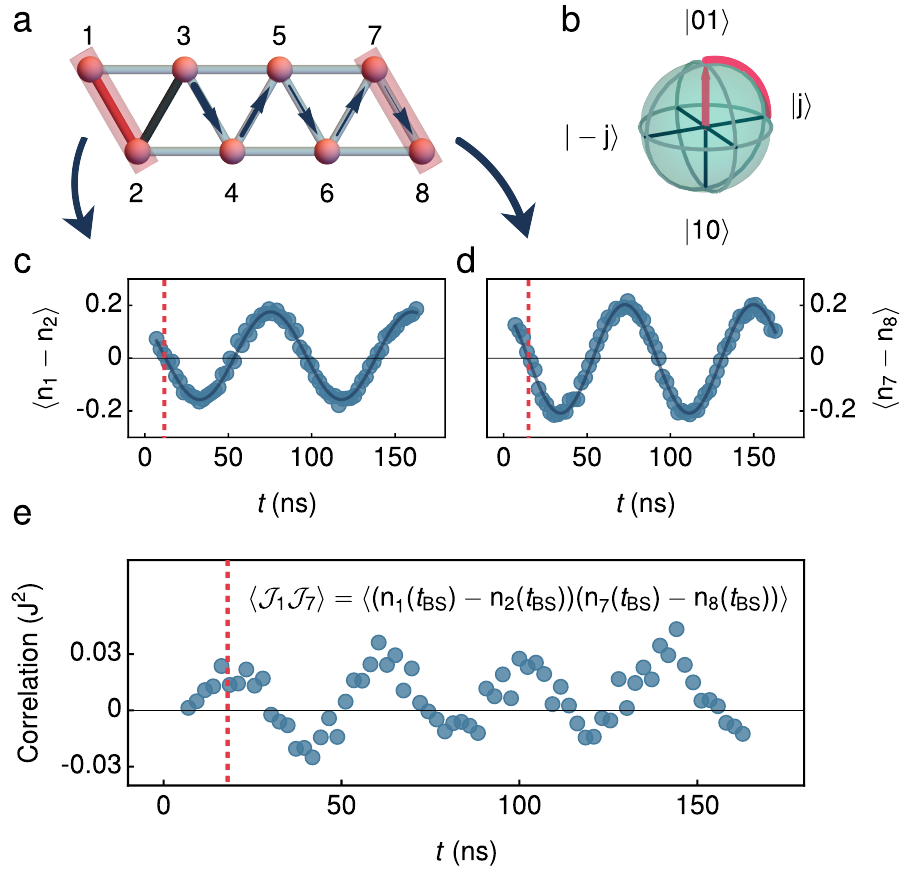}
    \caption{Current-current correlations between rungs of the triangular ladder. \textbf{(a)} Schematic of the lattice showing pairs of qubits undergoing beamsplitter rotation in order to perform current measurement. The arrows show the measured correlations with the rung shaded in red. \textbf{(b)} Bloch-sphere representation of the beamsplitter rotation from the current eigenstates $\ket{\pm j}$ to the readout basis. \textbf{(c)} Current measurement trace between qubits $1$ and $2$ as the beamsplitter interaction time is varied. The current value is extracted at the beamsplitter time $t^{\pdagger}_{\text{BS}}=\pi/4J$ shown by the dashed line. The solid line is a sinusoidal fit to the population oscillations. \textbf{(d)} Same as in (c) but for qubits $7$ and $8$. \textbf{(e)}. Current correlation between qubit pairs 1-2 and 7-8. Although the current on both rungs vanishes at the beamsplitter time, the current correlation has a nonzero value.}
    \label{fig:fig2}
\end{figure}

Viewing the two-qubit system in the single-particle manifold $\{\ket{01},\ket{10}\}$ as shown in Fig.~\ref{fig:fig2}a, we see that $\mathcal{J}_j$ and $H_{\text{\text{BS}}}$ act as effective $\sigma_y$ and $\sigma_x$ operations, respectively, on this degree of freedom. 
After applying an effective $\sigma_x$ to rotate the current basis to the measurement basis, we quickly detune all qubits to freeze dynamics and perform readout.

Upon quasiadiabatically preparing the ground state of the triangular ladder, we find that the rung currents vanish everywhere, as evidenced by the population differences vanishing at the beamsplitter time. Since the Hamiltonian is symmetric under time-reversal, for every state with current $\langle \mathcal{J}_j \rangle$ flowing from qubits $j$ to $j+1$, there is a degenerate state with current $-\langle \mathcal{J}_j \rangle$ flowing in the opposite direction. Without breaking time-reversal symmetry, the two current outcomes are equally likely, so after many iterations, the currents average to zero. Importantly however, the current-current correlation function,
\begin{equation}
G(i,j) = \langle \mathcal{J}_i \mathcal{J}_j\rangle - \langle \mathcal{J}_i \rangle \langle \mathcal{J}_j \rangle, 
\end{equation}
which can be further expanded into qubit-qubit population correlations using Eq.~\eqref{eqn:current},
can still be nonzero. 

Figure~\ref{fig:fig2} demonstrates the measurement of a nonzero current correlation between distant rungs, despite measuring zero average current, for $J_{\parallel}/J = -1$. First, we conduct the experimental procedure and read out a single pair of qubits, for example, qubits 1 and 2 in Fig.~\ref{fig:fig2}a. Since time evolution under $H_{\text{\text{BS}}}$ implements an effective $\mathrm{e}^{-\mathrm{i} Jt\sigma^{\pdagger}_x}$, it is necessary to calibrate $t_\text{BS}$ such that $\mathrm{e}^{-\mathrm{i} t_{\text{BS}} H_{\text{BS}}/\hbar} = \sqrt{\mathrm{i}\text{SWAP}}$, as seen in Fig.~\ref{fig:fig2}b,d. 
We elaborate further on the subtleties in calibrating the beamsplitter time from qubit swaps in the Supplemental Information (SI). The dashed lines in Figs.~\ref{fig:fig2}c,d indicate the fitted beamsplitter evolution time.

Having performed this calibration, we measure the correlations of currents between different pairs of qubits. For instance, the correlation between qubit pairs (1,2) and (7,8) is shown in Fig.~\ref{fig:fig2}e: while the dashed line corresponds to zero current, we extract a nonzero current correlation. However, if rungs $i$ and $j$ share a qubit, $(\mathcal J_i \mathcal J_j)^\dagger \neq \mathcal J_i \mathcal J_j$, and hence  $\langle \mathcal J_i \mathcal J_j \rangle$ is not measurable. There are six such pairs of rungs which we exclude, leaving $\binom{7}{2} - 6= 15$ combinations of rungs for which we repeat the above experiment.

\section*{Chiral current correlations}

\begin{figure*}[t]
    \includegraphics[width=\linewidth]{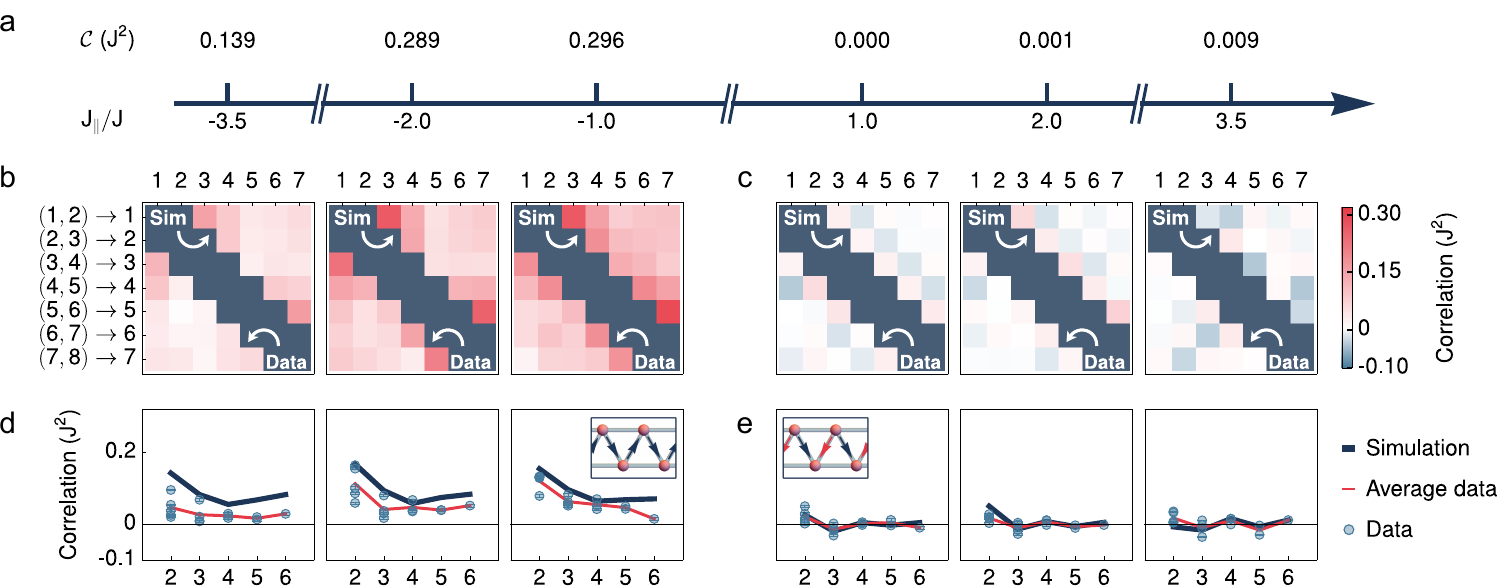}
    
    \caption{Chiral and Meissner superfluids on the 8-qubit triangle ladder. \textbf{(a)} Total chiral-current order parameter for various values of $J_{\parallel}/J$. Negative (positive) ratios correspond to a $\pi$-flux ($0$-flux) synthetic magnetic field. The values of $J_{\parallel}/J$ for each plot are (from left to right) $-3.56$, $-2.02$, $-1.22$, $0.98$, $1.96$, and $3.53$. \textbf{(b)} Current correlation matrices for the $\pi$-flux coupling ratios. Rung $j$ links qubits $(j,j+1)$. Experimental data are shown on the lower left and simulation on the upper right of the correlation matrix. \textbf{(c)} Current correlation matrices for the $0$-flux coupling ratios. \textbf{(d)} The same current correlations as in part (b) plotted against distance between the two rungs for the $\pi$-flux coupling ratios. The inset depicts a schematic of the rung currents that produce the positive current-current correlations observed here. \textbf{(e)} The same current correlations as in part (c) plotted against distance between the two rungs for the $0$-flux coupling ratios. The inset depicts a schematic of the rung currents that produce the anticorrelated current-current correlations observed here.}
    \label{fig:fig3}
\end{figure*}

\begin{figure*}[t]
    \includegraphics[width=\textwidth]{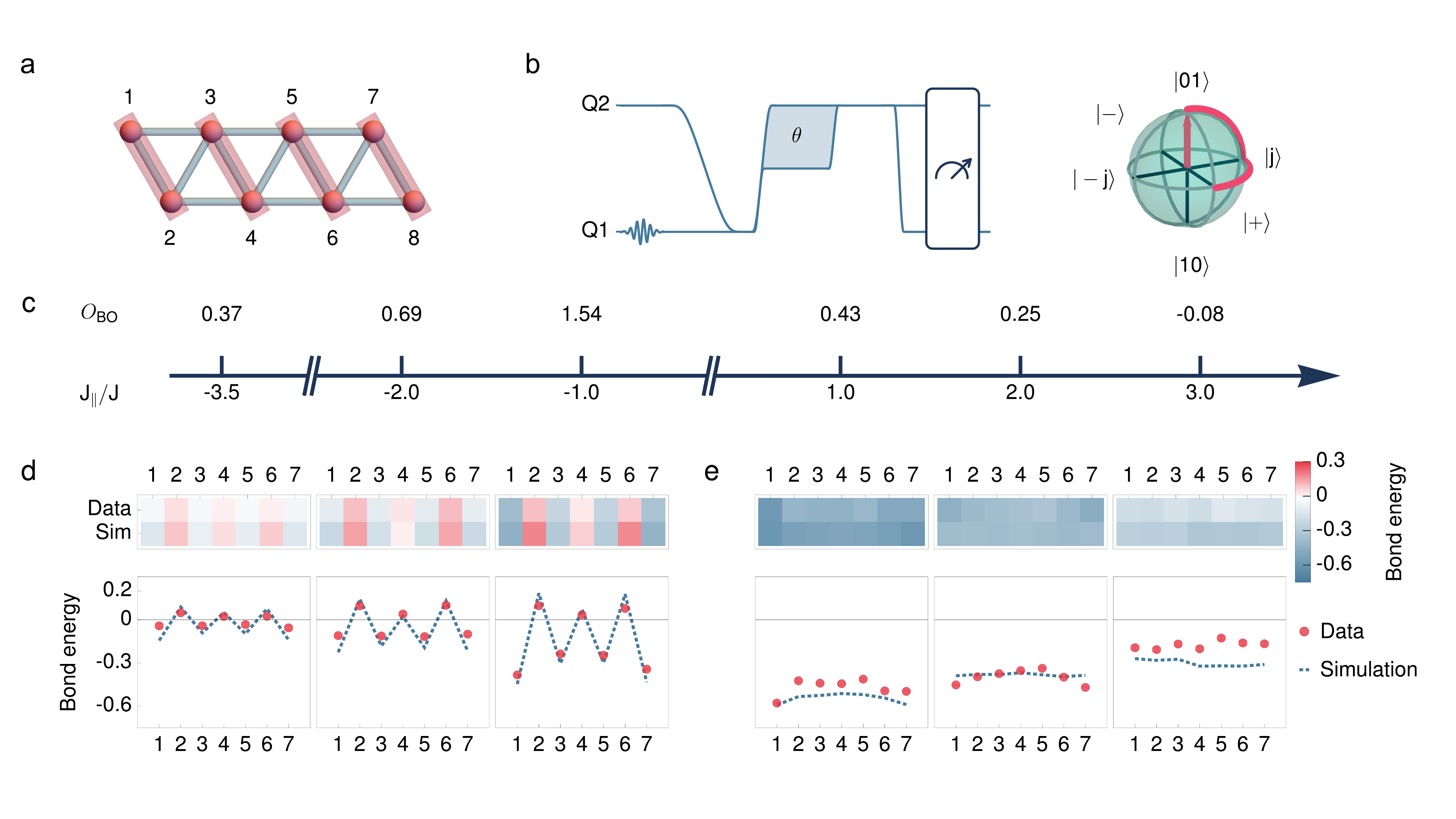}
    
    \caption{Bond order insulator in the 8-qubit triangle ladder. \textbf{(a)} Schematic of the bond kinetic energy ordering on the lattice. \textbf{(b)} Pulse diagram depicting the bond kinetic energy measurement. Here, one qubit acquires a phase relative to the other in order to rotate the kinetic energy ($\hat{x}$) axis to the current ($\hat{y}$) axis, before undergoing the beamsplitter interaction to rotate to the measurement ($\hat{z}$) axis. \textbf{(c)} Total bond order parameter for various values of $J_{\parallel}/J$: from left to right, $-3.56$, $-2.02$, $-1.22$, $0.98$, $2.04$, and $2.85$. \textbf{(d)} Bond kinetic energies of each rung for the $\pi$-flux coupling ratios. Experimental data (simulation) is shown on the upper (lower) row of the color plots. The dashed lines indicate the alternating trend in bond kinetic energies seen numerically. \textbf{(e)} Bond kinetic energy of each rung for the $0$-flux coupling ratios. Here, the experimental (red circles) and numerical (dashed lines) results highlight that the bond kinetic energies maintain a uniform sign throughout, without the alternation seen previously.
    }
    \label{fig:fig4}
\end{figure*}

Using this calibrated measurement procedure, we can now extract \textit{global} information on a many-body state of the triangular lattice, including the spatial texture of correlations. The color plots in Figs.~\ref{fig:fig3}b,c display the rung correlation matrix for each ratio $J_{\parallel}/J$. To emphasize the structure of these correlations, the same data are replotted in Figs.~\ref{fig:fig3}d,e as a function of the distance between the two rungs.

In the gapless quasi-1D superfluid regime, current correlations are generically expected to decay as a power law with distance~\cite{dhar2013chiral}. Accordingly, we indeed observe that the correlations generally decrease as the rung separation grows, for all measurement points. However, the plots in Figs.~\ref{fig:fig3}d,e show a slight increase in the most distant correlations relative to the monotonic trend. We attribute this behavior to finite-size and edge effects, since the largest separation corresponds to both rungs being on the boundary. Supporting this interpretation, correlations between rungs closer to the center of the lattice are systematically smaller at comparable separations. This position dependence manifests as a spread in the correlation values for a given separation in Figs.~\ref{fig:fig3}d,e, where we also plot the average correlation for each distance to compare with simulations. Both the decay of correlations and the edge effects are captured in these numerics, which we discuss further in the SI.

We first examine the correlations with an applied synthetic magnetic field corresponding to $\pi$ flux. As shown by the plots in Fig.~\ref{fig:fig3}b,d, when $J_{\parallel}/J < 0$, the current correlations are all positive. Given our sign convention in the definitions of currents, a current flowing from qubit 1 to qubit 2 is positively correlated with a current flowing from qubit 3 to qubit 4. This pattern is consistent with long-range chiral current order throughout the lattice~\cite{nishiyama2000finite, greschner2013ultracold}, since the currents flow with the same handedness around each similarly oriented plaquette.

The currents themselves average to zero across experimental iterations, implying that the $\pi$-flux ground state is a superposition of two superfluids with equal amplitude but opposite chirality, which are time-reversed copies of each other. 
Measuring a local current operator on this extended ground state projects the system into one of the two states with well-defined chiral currents, and this spontaneously breaks a discrete $\mathbb{Z}_2$ parity symmetry.  The nonzero chiral current-current correlations are characteristic of the chiral superfluid (CSF) phase, in which we have finite currents on the rungs that alternate sign from rung to rung (giving opposite circulations on adjacent plaquettes), together with leg currents that flow coherently along each leg.  These linked patterns of rung and leg currents correspond to circulating flow around elementary plaquettes and distinguish the CSF from nonchiral (for example Meissner-like or bond-ordered) phases~\cite{giamarchi,dhar2013chiral}.

In contrast to the $\pi$-flux case discussed above, we notice a stark difference in the correlations in the absence of a synthetic magnetic field. In this case, we no longer expect states to have chiral properties and the plots in Fig.~\ref{fig:fig3}c,e for $J_{\parallel}/J>0$ indeed demonstrate the breakdown of the chiral order. In addition to the smaller magnitude, the correlations also tend to alternate sign with increasing distance. In this regime, rungs exhibit anticorrelated currents, indicating the absence of a uniform chirality. This is characteristic of a Meissner superfluid (MSF)~\cite{orignac2001meissner, cha2011two}, which ideally has no currents along the rungs and no chirality around plaquettes; the nonzero correlations we measure here arise from having a finite system.

We can define the chiral-current order parameter as the sum over distance-averaged correlations:
\begin{equation}
    \mathcal{C} = \sum_d \frac{1}{N-d} \sum_j[G(j,j+d)].
\end{equation}
We report $\mathcal{C}$ for each coupling ratio along the axis in Fig.~\ref{fig:fig3}a. The order parameter measurement reveals that $\mathcal C$ differs qualitatively between the $\pi$-flux and $0$-flux cases: the measured values are orders of magnitude smaller in the latter case than the former, which is consistent with our expectation that $\mathcal C$ should vanish for a MSF but remain nonzero for a CSF.

\section*{Bond order insulator}

While current correlations provide a direct probe of chiral transport and distinguish key features between the $0$- and $\pi$-flux regimes, more generally, the complex nearest-neighbor correlator $\langle a_j^\dagger a_{j+1}^{\pdagger}\rangle$ contains both imaginary and real components, which encode distinct physical information. In contrast to the imaginary part discussed above (which is proportional to the particle-number current), the real part,
$\mathcal{O}^{\pdagger}_j$\,$=$\,$2\,\mathrm{Re} [\langle a_j^\dagger a_{j+1}^{\pdagger}\rangle],
$
corresponds to the expectation value of the kinetic energy operator associated with tunneling on a given link $(j,j+1)$. This quantity characterizes the degree to which particles are coherently delocalized across neighboring sites and thus provides a natural probe of bond-energy ordering in the many-body ground state. 

Experimentally, we access $\mathcal{O}_j$ using a modified version of the beamsplitter measurement used for current detection, as seen in Fig.~\ref{fig:fig4}b. Prior to the beamsplitter pulse, we allow the system to evolve under an idling Hamiltonian
$H_{\text{idle},j}$\,$=$\,$\hbar \Delta_j \big (a_j^\dagger a_j^{\pdagger}$\,$-$\,$a_{j+1}^\dagger a_{j+1}^{\pdagger}\big)$
implemented by detuning the two qubits forming the bond such that $|\Delta_j|\gg J$. Choosing an evolution time $t_{\text{idle}}=\pi/4\Delta_j$ rotates the $\hat{x}$-axis to $\hat{y}$ on the Bloch sphere during idling, and then $\hat{y}$ to $\hat{z}$ during the beamsplitter. This, as before, effectively maps the real part of $\langle a_j^\dagger a_{j+1}^{\pdagger}\rangle$ onto a population imbalance in the computational basis.

Figure~\ref{fig:fig4} shows the measured bond kinetic energies $\mathcal{O}_j$ for each coupling ratio. As $|J_{\parallel}/J|$ increases, the magnitude of the bond energies on the rungs decreases, since the bosons preferentially delocalize along the legs when leg hopping dominates. In the presence of a $\pi$ synthetic flux (Fig.~\ref{fig:fig4}d), however, the rung bond energies exhibit a qualitatively distinct pattern: their signs alternate from rung to rung. Contrarily, in the zero-flux case (Fig.~\ref{fig:fig4}e) the bond energies maintain a uniform sign across the ladder.

To quantify this behavior, we define the bond order parameter~\cite{giamarchi}
\begin{equation}
O^{\pdagger}_{\text{BO}} = \sum_j^{N-1} (-1)^j \left(\langle a_j^\dagger a_{j+1}^{\pdagger}\rangle + \langle a_{j+1}^\dagger a_j^{\pdagger}\rangle \right),
\end{equation}
which measures the staggered component of the bond kinetic energy. A nonzero value of $O_{\text{BO}}$ indicates that the kinetic energy alternates between neighboring bonds, breaking lattice translational symmetry. In our measurements, the order parameter is largest in the $\pi$-flux configuration near $J_{\parallel}/J$\,$\approx$\,$-1$, where the rung bond energies have large magnitudes and alternate in sign. As $|J_{\parallel}$\,$/$\,$J|$ increases further, the magnitude of $O_{\text{BO}}$ decreases because the rung bond energies themselves become small, whereas in the absence of a synthetic field, the order parameter remains suppressed because the bonds no longer alternate.

This staggered pattern of bond energies is the defining signature of the bond order insulator (BOI) phase predicted for frustrated bosonic ladders in a magnetic field~\cite{giamarchi}. This bond ordering arises from the combination of strong interactions, geometric frustration, and flux, which conspire to favor a ground state with modulated kinetic energy density rather than uniform superfluid coherence.
Intuitively, repulsive onsite interactions suppress macroscopic occupation of a single site and penalize delocalization patterns that place large density fluctuations on individual sites.  Faced with a frustrated kinetic term that weakens the homogeneous delocalization advantage, the interacting system can instead optimize its energy by reorganizing the pattern of tunneling: concentrating kinetic energy density on a subset of bonds while suppressing it on others.  The result is a spatially modulated expectation value of the bond kinetic operator $(\mathcal{O}_j)$, manifesting as alternating strong and weak bonds.

In the many-body phase diagram, the gapped BOI lies next to the gapless chiral superfluid regime in parameter space~\cite{giamarchi}. This proximity implies an enhanced susceptibility toward bond ordering as $|J_{\parallel}/J|$ is tuned towards the critical point. Since our device operates near this phase boundary (albeit on the CSF side) and because the ideally sharp phase transition is dulled by the finite size of our device, we measure a nonzero $ O_{\text{BO}}$ signal, reflecting the incipient instability towards bond ordering and the phase transition between these competing many-body states.

\section*{Discussion and Outlook}

In these experiments, we study a superconducting-qubit device that realizes an interacting Bose-Hubbard model---a paradigmatic model for understanding strongly correlated quantum matter---on a triangular ladder at half filling. Exploiting the tunability of transmon couplers, we implement a synthetic magnetic flux of $0$ or $\pi$ per plaquette and tune the ratio between rung and leg tunneling amplitudes to explore a rich ground-state structure predicted for frustrated bosonic ladders~\cite{giamarchi}. 
We prepare and characterize multiple many-body states, including chiral superfluids, which host staggered loop currents that spontaneously break time-reversal symmetry,  and Meissner superfluids with vanishing rung currents. 
These phases are distinguished by measuring link-current operators and correlators thereof, which we access experimentally through a calibrated mapping from qubit population dynamics to the expectation values of currents. 
Using the same measurement framework, we also identify signatures of incipient bond ordering, reflected in a pattern of staggered bond kinetic energies along the rungs of the ladder.

Chiral superfluids and related current-carrying phases have been widely proposed~\cite{ barbiero2023frustrated} and, more recently, experimentally explored~\cite{li2023observation} in quantum simulators based on ultracold atoms. In the context of such atomic systems, most studies on the triangular Bose-Hubbard ladder Hamiltonian have focused on unit filling~\cite{ greschner2013ultracold, zaletel2014chiral, romen2018chiral} or low densities~\cite{ greschner2019interacting}. Superconducting circuits provide a complementary approach with several practical advantages. 
In particular, our setup of transmon qubits facilitates deterministic high-fidelity preparation of many-body states with a fixed number of excitations, whereas cold-atom experiments typically rely on statistical loading of lattice sites. 
This capability enables us to initialize arbitrary product states in a chosen particle-number sector and subsequently ramp onto the target Hamiltonian. 
In addition, superconducting platforms offer rapid experimental repetition rates and precise control over individual couplings and site parameters, allowing Hamiltonian parameters to be engineered both in device design and through in situ tuning. 
These features provide a versatile environment for exploring frustrated Bose-Hubbard physics with direct microscopic control that may be more challenging to achieve in optical lattices.

More broadly, the techniques demonstrated here provide a useful toolset for the analog quantum simulation of lattice boson models and their strongly correlated phases. 
Extending these experiments to wider parameter ranges is a natural next step for systematically exploring the full phase diagram, including the critical regimes separating Meissner and chiral superfluids, and bond-ordered phases. 
This opens the door to addressing important theoretical questions on quantum criticality: measuring critical exponents and related universal properties would provide direct experimental tests of field-theoretic and numerical predictions for quantum phase transitions in frustrated bosonic systems~\cite{giamarchi2003quantum,dhar2013chiral}.
Going beyond equilibrium ground states, our platform further provides a window into the fundamentally nonequilibrium quantum dynamics of state preparation. 
For example, the controlled parameter ramps through the Mott-insulator--superfluid quantum phase transition offer a natural testbed to probe the dynamical emergence of coherence and superfluid order~\cite{braun2015emergence}. It would also be interesting to understand the dynamical aspects of $\mathbb{Z}_2$ symmetry breaking and domain formation in chiral sectors~\cite{shimizu2018, coarsening2024}.
Moreover, the control and tunability afforded by superconducting circuits makes the system well suited for extracting more refined metrics, such as multipartite entanglement~\cite{singha2022genuine}, and for investigating finite-temperature properties of interacting lattice models~\cite{buser2019}, regimes that remain challenging to access with classical numerical methods.

On the experimental front, one can envision several improvements to further enhance the versatility of our platform. 
At present, qubit frequencies can be tuned rapidly during an experimental sequence, whereas coupler parameters are adjusted more slowly between runs. 
Expanding control-hardware capabilities such that couplers can be tuned dynamically on experimental timescales could permit the design of more flexible ramp protocols and enable studies of adiabatic state preparation and thermalization dynamics~\cite{Andersen2025ThermalizationCriticality}. 
Increasing the accessible range of coupler tunability---particularly the ability to tune the effective coupling through zero---would also simplify state initialization and improve readout fidelity for neighboring qubits. 
Such tunability would provide insights at intermediate coupling ratios between the chiral and Meissner regimes, where phase transitions are expected, as well as deeper access to the parameter region associated with bond-ordered phases. 
Finally, Floquet modulation of qubit or coupler parameters could be used to engineer synthetic gauge fields beyond the static $0$- and $\pi$-flux configurations studied here~\cite{Roushan2017Chiral}. 
Access to continuously tunable flux would allow explicit breaking of time-reversal symmetry and thus, shed new light on additional phases, including biased chiral superfluids with nonzero average current~\cite{giamarchi}.

\section*{Acknowledgments}
We thank Faranak Bahrami, R. N. Bhatt, Eugene Demler, Lev Krayzman, and Basil  Smitham for useful discussions, and Sho Uemura for assistance with QICK software development. We acknowledge support primarily from the NSF Quantum Leap Challenge Institute for Robust Quantum Simulation 2120757, with additional support for theoretical work from U.S. Department of Energy, Office of Science, National Quantum Information Science Research Centers, Co-design Center for Quantum Advantage ($\text{C}^2\text{QA}$) under Contract No. DESC0012704. R.S. was partially supported by the Princeton Quantum Initiative Fellowship. This research made use of the Micro and Nano Fabrication Center at Princeton University.

Princeton University Professor Andrew Houck is also a consultant for Quantum Circuits Incorporated (QCI). Due to his income
from QCI, Princeton University has a management plan in place to mitigate a potential conflict of interest that could affect the
design, conduct and reporting of this research.

\bibliographystyle{apsrev4-2_custom} 
\bibliography{refs}

\section*{Methods}

\subsection*{Device fabrication}

The eight-qubit device in this work consists of a \SI{200}{\nm}-thick layer of tantalum deposited on a \textit{c}-axis HEMEX sapphire substrate from Crystal Systems before undergoing the cleaning process outlined in Ref.~\cite{BahramiVortices}. Afterwards, the base layer pattern is defined with optical lithography and etched with a chlorine-based dry etch. Josephson junctions are patterned with electron-beam lithography to create a suspended resist bridge \cite{Dolan}. The Al-$\text{AlO}_x$-Al layers are deposited using using double-angle electron-beam evaporation in a custom UHV-chamber Plassys \cite{bland2025millisecond}. The first layer is \SI{20}{\nm} thick and the $\text{AlO}_x$ is grown for 40 minutes in pure oxygen at $50~\mathrm{mbar}$. The second layer is $\SI{70}{nm}$ thick. Afterwards, aluminum airbridges are patterned with two optical lithography steps following the process in Ref.~\cite{ChenAirbridge}, and deposited by similar means in a Plassys MEB550S system. Airbridges are use to short ground planes on opposite sides of coplanar waveguides as well as to connect qubit and coupler flux lines on opposite sides of the transmission line. Aluminum wirebonds are used to connect signal lines to bond pads on flux lines and the transmission line. The device is mounted in the Scalinq LINQER package.

\subsection*{Wiring and setup}
Devices were measured in a Bluefors dilution refrigerator with a base temperature of $10~\mathrm{mK}$.
Time-domain control and measurements were performed using the Xilinx RFSoC ZCU216
with experiments written in QICK~\cite{Stefanazzi_2022},
including resonator and qubit drives, qubit fast-flux
pulses, and simultaneous multiple-cavity
readout. DC-flux offsets for qubits and couplers were controlled with
Qblox D5a voltage sources, combined with
a fast-flux line in the mixing chamber, and sent to the
qubit via on-chip flux lines.
The qubits were dispersively read out and driven
through individual quarter-wave resonators inductively
coupled to a single transmission line. Resonator readout
tones and qubit drive pulses were combined using a 6dB directional coupler
at room temperature. The output signal was amplified
using a traveling-wave parametric amplifier (TWPA) at 10mK, a high-electron-mobility transistor (HEMT) amplifier at 4K, and two amplifiers at room temperature. A Windfreak SynthHDv2 RF source was used
to provide the pump drive for the TWPA.
The full wiring setup can be found in Fig.~S1 of the Supplementary Information.

\subsection*{Experimental sequence}
For all experiments, DC-flux voltages are set so that all eight qubit frequencies are near the midpoint of their respective ranges of tunability. This enabled frequency tuning across the entire spectrum with solely fast-flux pulse control. DC-flux voltages for the six couplers are set to
produce the desired coupling strength $J_{||}$ when the qubits are at the lattice resonance frequency.
State initialization, time evolution, and readout are
then performed using fast-flux pulses that control the
qubit frequencies within \SI{0.3}{\ns}.
State preparation entails tuning the frequencies of qubits 1, 4, 8, and 5 to be evenly spaced by \SI{50}{\MHz} and driving each of these four qubits to the excited state. All other qubits are negatively detuned by \SI{150}{\MHz} for this step. Qubits 1, 4, 8, and 5 are then quickly tuned to the lattice frequency ($\omega_\text{lattice}$), and
qubits 2, 3, 6, and 7 are slowly ramped onto the same resonance frequency over \SI{300}{\ns}. During the measurement step, qubit pairs are tuned onto resonance with each other, with different resonant frequencies for each pair, depending
on which qubits are designated to undergo the beamsplitter operation;
qubits not involved are far detuned during the beamsplitter time evolution.
Finally, qubits are detuned by at least 20$J$ from other qubits for simultaneous single-shot readout of all eight qubits.

Readout for the current correlation and bond kinetic energy data consists of collecting 100,000 shots of simultaneous eight-qubit single-shot measurements. Readout parameters, chosen to optimize discrimination between the ground and first excited states, are determined before every run by collecting 8000 shots of the readout calibration experiment.

\subsection*{Tuning qubits onto resonance}
We calibrate the qubit frequencies after the adiabatic ramp to ensure that all qubits are brought on resonance. To this end, we extract the frequency of each qubit from a Ramsey experiment with the drive frequency $\omega_\text{drive} = \omega_\text{lattice}$ while tuning the fast-flux pulse amplitude. This, combined with accounting for flux crosstalk, allows us to fine-tune all qubits to be resonant at $\omega_\text{lattice}$.

\subsection*{Hamiltonian and Lindbladian simulations}

All parameters used in the Hamiltonian and Lindbladian simulations of the dynamics are independently measured.
Coupling strengths are measured by extracting the rate of resonant exchange of a single excitation
between all pairs of qubits. $T_1$ and $T_{2R}$
decoherence times are measured at $\omega_\text{lattice}$ while all other qubits are far detuned. Anharmonicities are measured by calibrating the frequency of an $\omega_{12}$ pulse after preparing
a qubit in $\ket{1}$.

\newpage


\section*{Supplementary material}

# Supplementary Information for "Chiral and bond-ordered phases in a triangular-ladder superconducting-qubit quantum simulator"

## S1. DEVICE PARAMETERS

| Qubit | Q1 | Q2 | Q3 | Q4 | Q5 | Q6 | Q7 | Q8 |
|---|---|---|---|---|---|---|---|---|
| $\alpha/2\pi$ (MHz) | 188 | 187 | 184 | 187 | 183 | 186 | 187 | 187 |
| $\omega_{q,\mathrm{bare,min}}/2\pi$ (MHz) | 3524 | 3516 | 3473 | 3517 | 3512 | 3513 | 3448 | 3512 |
| $\omega_{q,\mathrm{bare,max}}/2\pi$ (MHz) | 4409 | 4418 | 4378 | 4434 | 4402 | 4380 | 4324 | 4417 |
| $\kappa/2\pi$ (MHz) | 0.59 | 1.05 | 0.93 | 1.07 | 1.06 | 0.88 | 0.73 | 0.87 |
| $\chi/2\pi$ (MHz) | 0.53 | 0.41 | 1.10 | 0.56 | 0.41 | 0.36 | 0.68 | 0.52 |
| $\omega_r/2\pi$ (MHz) | 7121 | 7077 | 7511 | 7568 | 7363 | 7441 | 7255 | 7309 |
| | | | | | | | | |
| $(J_{||}/J = -3.5)$ $T_1$ (µs) | 37.5 | 38.1 | 33.5 | 24.5 | 16.1 | 19.4 | 21.4 | 28.8 |
| $(J_{||}/J = -2.0)$ $T_1$ (µs) | 39.3 | 51.5 | 39.3 | 30.5 | 19.2 | 32.8 | 24.5 | 33.6 |
| $(J_{||}/J = -1.0)$ $T_1$ (µs) | 31.1 | 42.7 | 26.6 | 16.9 | 19.4 | 23.3 | 23.2 | 33.6 |
| $(J_{||}/J = +1.0)$ $T_1$ (µs) | 32.8 | 45.2 | 39.5 | 37.8 | 15.7 | 28.5 | 23.7 | 24.1 |
| $(J_{||}/J = +2.0)$ $T_1$ (µs) | 25.6 | 33.9 | 26.1 | 28.4 | 15.6 | 23.9 | 21.4 | 30.0 |
| $(J_{||}/J = +3.5)$ $T_1$ (µs) | 16.5 | 19.3 | 15.8 | 21.0 | 9.4 | 14.9 | 14.7 | 16.8 |
| | | | | | | | | |
| $(J_{||}/J = -3.5)$ $T_{2R}$ (µs) | 1.47 | 0.53 | 1.54 | 0.42 | 2.40 | 2.04 | 1.37 | 1.62 |
| $(J_{||}/J = -2.0)$ $T_{2R}$ (µs) | 1.77 | 0.48 | 2.60 | 0.51 | 3.18 | 2.62 | 2.21 | 2.23 |
| $(J_{||}/J = -1.0)$ $T_{2R}$ (µs) | 1.04 | 0.96 | 0.53 | 1.00 | 0.65 | 0.59 | 1.48 | 0.84 |
| $(J_{||}/J = +1.0)$ $T_{2R}$ (µs) | 2.19 | 0.74 | 2.60 | 1.49 | 2.08 | 2.52 | 2.33 | 1.93 |
| $(J_{||}/J = +2.0)$ $T_{2R}$ (µs) | 3.31 | 1.07 | 4.15 | 1.00 | 5.12 | 3.74 | 5.69 | 3.50 |
| $(J_{||}/J = +3.5)$ $T_{2R}$ (µs) | 2.09 | 0.22 | 3.40 | 0.27 | 3.53 | 2.94 | 5.45 | 2.50 |

TABLE S1. A table of measured device parameters for the 8-qubit triangular ladder. $\alpha = \omega_{01} - \omega_{12}$ is the qubit anharmonicity. $\omega_{q,\mathrm{bare,min}}$ and $\omega_{q,\mathrm{bare,max}}$ are the minimum and maximum tunable bare qubit frequencies, respectively. $\kappa$ is the resonator linewidth and photon loss rate. $\chi$ is the change in the resonator frequency when the qubit is in its first excited state versus its ground state. $\omega_r$ is the resonator frequency when the qubit is in its ground state. $\omega_r$ varies over a range of 1–2 MHz depending on qubit frequency due to the dispersive resonator-qubit coupling. $T_1$ and $T_{2R}$ are the qubit decay and Ramsey dephasing times, respectively, between the ground and first excited states, all measured at the resonance point of 3800 MHz for $J_{||}/J \in \{-3.5, -2.0, -1.0, +1.0\}$ or 4295 MHz for $J_{||}/J \in \{+2.0, +3.5\}$.

## S2. WIRING AND SETUP

The wiring and control setup is shown in Fig. S1. For arbitrary pulse generation, pulse scheduling, and signal digitization, we use the Xilinx RFSoC ZCU216 with experiments written using the QICK library [1]. This setup enables synthesis of qubit drives, readout probes, and frequency control of qubits and couplers, which we refer to as "fast flux". We include bias tees to combine fast-flux lines with DC lines in the mixing chamber; these have been modified by replacing the capacitor with a short.

## S3. FREQUENCY CALIBRATION

For each of our eight qubits, we measure the frequency by performing two-tone spectroscopy while varying DC voltage on the corresponding flux-bias line. We fit the flux dependence of the frequency to an asymmetric-junction tunable-transmon model [2, 3]. In order to model the coupler's frequency spectrum, we rely on tracking the avoided crossings between the coupler and its neighboring qubits, as shown in Fig. S3.

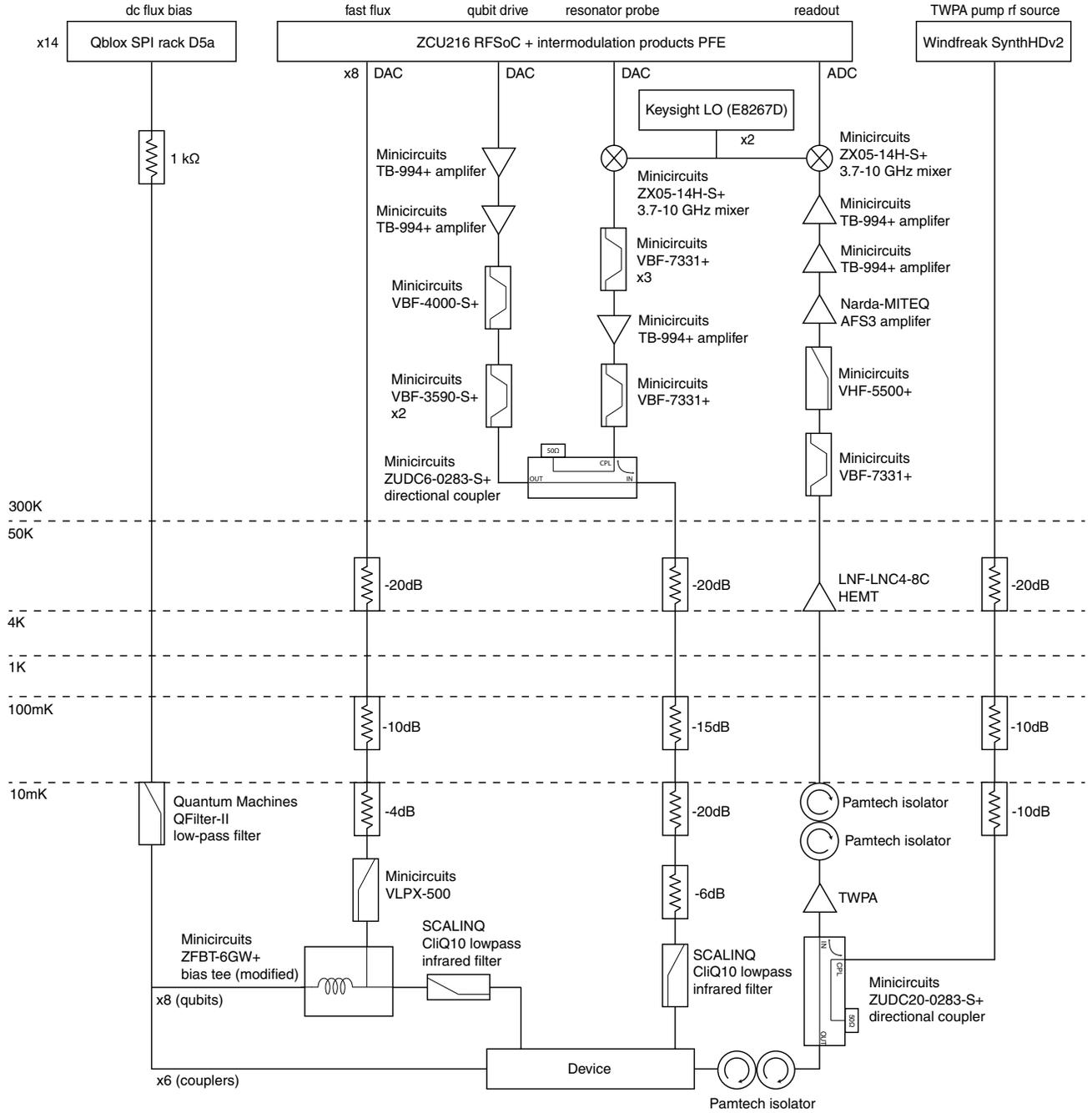

FIG. S1. Control and measurement wiring setup.

After calibrating the qubits and couplers, we measure frequency shifts arising from applied biases on other flux lines in order to characterize the flux crosstalk matrix [4, 5]. For each qubit, we generate random sets of bias voltages on the other flux lines and measure the difference in the target qubit's frequency from the expected frequency based on the tunable-transmon model. We use least-squares minimization to find a crosstalk matrix that best maps the randomly generated voltages to the corresponding measured frequency. We update the expected frequency based on the crosstalk matrix and repeat the process to further increase the precision of the crosstalk model.

The procedure for calibrating crosstalk on the couplers is slightly different but again relies on measuring avoided crossings with neighboring qubits. We pick the coupler frequency to be at the highest value in its range, which is identifiable by recognizing the symmetry between its avoided crossings with adjacent qubits. As before, we randomize

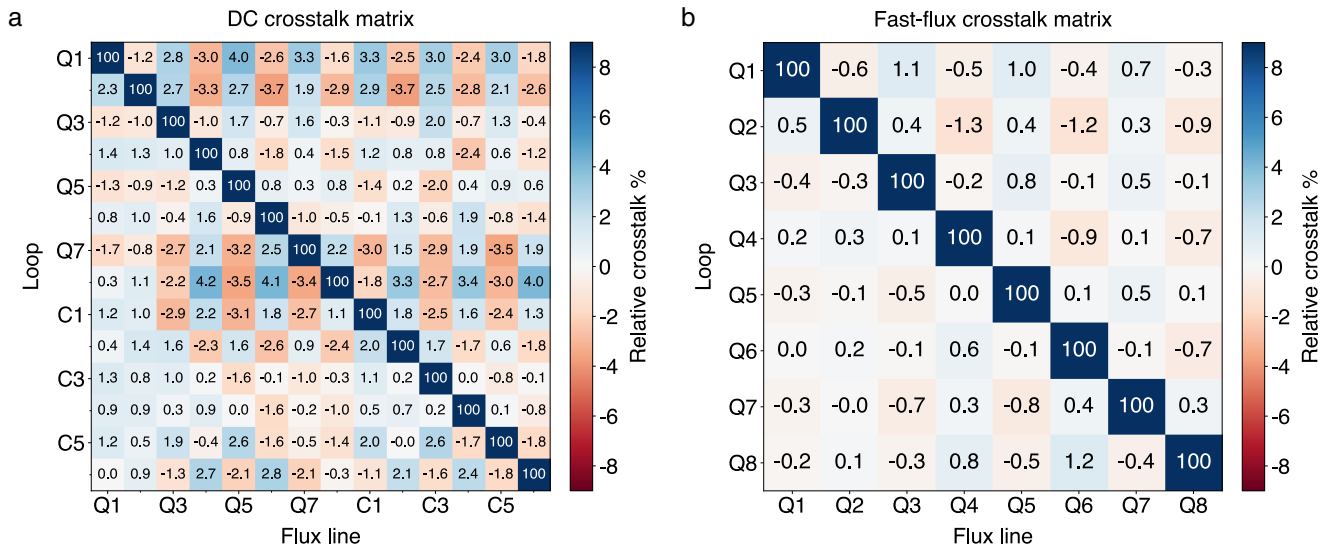

FIG. S2. Flux crosstalk matrices. **(a)** DC flux bias crosstalk matrix. The color indicates the relative crosstalk, or the amount of magnetic flux through a qubit or coupler's junction loop (row) originating from a different qubit or coupler's associated flux bias line (column). **(b)** Fast-flux crosstalk matrix.

the frequencies of all other qubits (and voltages for all other couplers), and measure the qubit-coupler avoided crossings. We use least-squares optimization to find the crosstalk matrix that best predicts the shift in the coupler's avoided crossing due to the bias voltages on all other flux lines.

To calibrate a crosstalk matrix for the fast-flux channels on the eight qubits, we perform the same procedure but replace the DC biases with varying-amplitude flat pulses from our RFSoC. In doing so, we obtain both the 14×14 DC crosstalk and the 8×8 fast-flux crosstalk matrices, both shown in Fig. S2.

## S4. CALIBRATION OF COUPLING STRENGTHS

Experimentally, we determine the coupling strengths by measuring the frequency of oscillations between the qubits. When nearly resonant, two qubits exchange a single excitation at a rate given by

$$\Omega_{\text{swap}} = 2\sqrt{J^2 + \Delta^2/4},\tag{S1}$$

where $J$ is the coupling strength and $\Delta$ is the detuning between the qubits. Thus, we measure particle swaps while varying the detuning to extract the coupling strength.

While qubits are directly coupled along the rungs of the ladder, the coupling between qubits on the legs can be modeled as a second-order hopping process mediated by the coupler mode, with an effective coupling given by [6],

$$g_{\text{eff}} \approx \frac{g_1 g_2}{\omega_c - \omega_q} \cdot \frac{2\omega_q^2}{\omega_c(\omega_q + \omega_c)} + g_{12}.\tag{S2}$$

Here, $g_1$ and $g_2$ are the couplings between the qubits and the couplers, $g_{12}$ is the direct coupling between the qubits, and $\omega_c$ and $\omega_q$ are frequencies of the coupler and (resonant) qubits, respectively, all of which are depicted in Fig. S3a. It is readily seen from the denominator of the first term that when $\omega_c > \omega_q$, $g_{\text{eff}} > 0$ (ignoring $g_{12}$ which is small in our device), and when $\omega_c < \omega_q$, $g_{\text{eff}} < 0$. Therefore, the sign of $g_{\text{eff}}$ is selected by whether the coupler frequency is above or below the qubit frequencies.

A consequence of the strong qubit-coupler couplings is a noticeable difference between the bare and dressed frequencies of the full 8-qubit and 6-coupler system. Therefore, when mapping qubit bias fluxes to frequencies, we account for frequency dressing by diagonalizing the subsystem consisting of a qubit and its neighboring couplers. While determining the bias fluxes needed to tune all qubits into resonance, we minimize the effects of such dressing by performing Ramsey experiments to measure individual qubit frequencies. In this procedure, we drive the $\pi/2$ pulses of the Ramsey sequence at the target lattice frequency. The optimal flux-pulse amplitude is then identified as the point where the qubit-drive detuning is minimized, which is indicated by the suppression of Ramsey fringes.

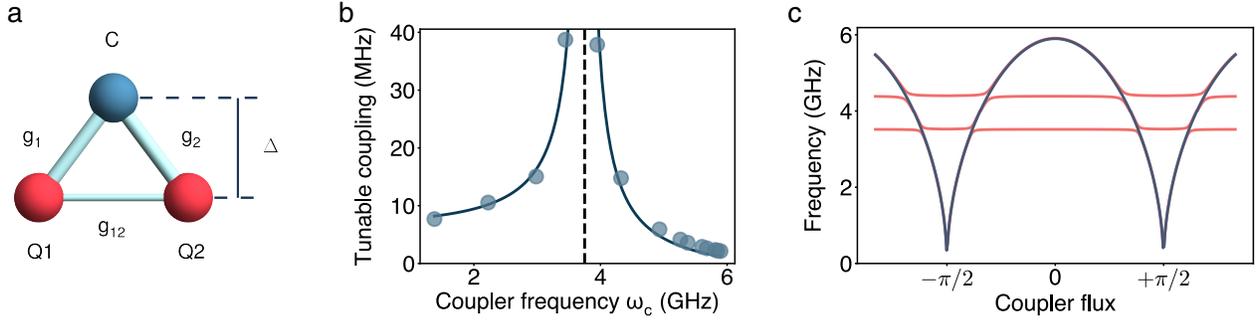

FIG. S3. Tunable leg couplings. **(a)** The qubit-coupler-qubit subsystem used to model the effective coupling strengths between qubits. **(b)** The magnitude of the effective coupling strength $|g_{\text{eff}}|$ as a function of the coupler frequency $\omega_c$ when the qubit frequency is set to $\omega_q = 3800\,\text{MHz}$ (dashed vertical line). Blue markers indicate the measured effective coupling strengths for various coupler frequencies. The dark blue curve plots the expression in Eq. (S2) using $g_{12}$ as free fit parameter. Values for $g_1$ and $g_2$ were acquired from the values of the energy gap measured at avoided crossings of the qubit and coupler modes. **(c)** We find the frequency spectrum of the couplers by performing two-tone spectroscopy on the hybridized qubit-coupler states near avoided crossings between a coupler and its two neighboring qubits, utilizing the qubit's readout hardware. The dark blue line is the fit to the bare coupler spectrum, and the light red lines are fits to the hybrid qubit-coupler eigenstates.

## S5. FLUX PULSE DISTORTION

We characterize the fast-flux pulses by measuring the qubit frequency at various time points in order to compensate for distortions from the ideal step pulse. We fit the frequency to a step pulse multiplied by an envelope containing three exponentially decaying functions [7, 8]:

$$f(t) = 1 + A_1 \exp(-t/\tau_1) + A_2 \exp(-t/\tau_2) \cos(\omega_2 t + \phi_2) + A_3 \exp(-t/\tau_3).$$ (S3)

We perform a $Z$-transform to find the analytic infinite-impulse response (IIR) filter of this function, and we apply the inverted filter to all waveforms sent through the corresponding DAC. The results in Fig. S4 demonstrate the difference in distortion with no calibration and after three iterations of calibrating the filter function.

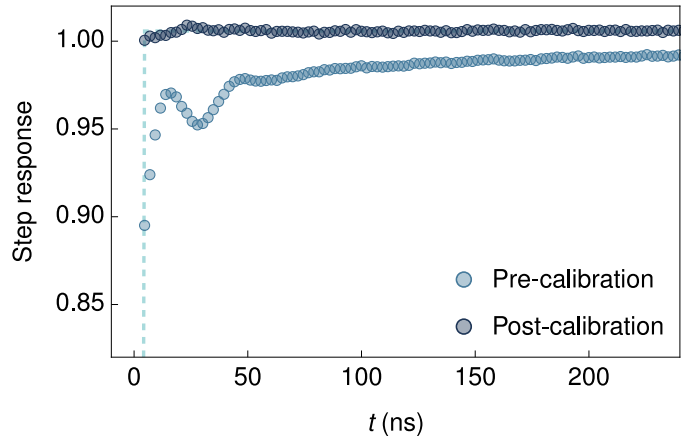

FIG. S4. Qubit frequency over time, acquired via two-tone spectroscopy, immediately after a flat step pulse is sent through the fast-flux line. Light blue shows the behavior before applying pulse predistortion, and dark blue shows the behavior after.

## S6. TIMING CALIBRATION FOR BEAMSPLITTER MEASUREMENTS

To perform the beamsplitter measurement, we detune the lattice into resonant qubit pairs separated from all other qubits by more than $20J$. Measuring both the real and imaginary parts of $\langle a_i^\dagger a_j \rangle$ requires precise control over the phase between the two qubits. No phase difference corresponds to a measurement of the current, while a phase difference of $\pi/2$ corresponds to a measurement of the bond kinetic energy. To this end, we add an intermediate step between the ramp and the beamsplitter interaction where the two qubits idle with some frequency difference, as shown in Fig S5a. Tuning the idle duration $t_{\mathrm{idle}}$ and frequency difference allows measurement of $\langle a_i^\dagger a_j \rangle$ at arbitrary complex phase.

We begin by characterizing the relative arrival times of fast-flux pulses applied to different qubits, in order to apply detuning step pulses simultaneously. Following the procedure in Ref. [5], we send a qubit drive pulse with some delay relative to a fast-flux pulse on the same qubit. When the pulses are aligned, the qubit frequency is resonant with the drive and we measure a high excited-state population. Otherwise, the qubit frequency is off-resonant from the drive and no signal is measured. Figure S5b reveals a maximum excited-state population of qubit 1 for a delay of around 80 ns. We repeat this experiment for all qubits to determine the relative delays required for simultaneous qubit fast-flux pulses.

With the pulse delays calibrated, we tune the ideal duration and frequency difference to determine the phase difference. Figure S5c shows the population difference between the two qubits after preparing the state $|10\rangle + |01\rangle$, applying the phase offset by idling, and measuring the beamsplitter population swaps. As we vary $t_{\mathrm{idle}}$ on the $y$-axis, we notice periodic patterns of high-contrast or no-contrast population swaps. No-contrast traces correspond to a total phase of $n\pi$ for integer $n$ (since $|10\rangle \pm |01\rangle$ is invariant under the swap Hamiltonian), and high-contrast traces correspond to a phase of $(n + \frac{1}{2})\pi$ (when the state rotates into $|10\rangle \pm i |01\rangle$). This calibration also absorbs residual phase offsets from asynchronous fast-flux pulses arising due to the finite waveform timing resolution (290 ps). After selecting the $t_{\mathrm{idle}}$ that minimizes swap contrast (for current measurement) or maximizes swap contrast (for bond kinetic energy), we perform another round of calibrations to fine-tune the frequency difference during idling since finer resolution is accessible in the fast-flux amplitude than in the waveform time delay.

Finally, after calibrating the idle duration and frequency difference, we fit the population swap trace to a sinusoid to extract the time that corresponds to a phase of $\pi/2$, representing the rotation on the Bloch sphere from the current ($\hat{y}$) axis to the measurement ($\hat{z}$) axis. Since the current measurement is calibrated to have minimal contrast of the two-qubit swaps, we instead ramp the two qubits to a slightly off-resonant point to prepare the state $\alpha |10\rangle + \sqrt{1 - \alpha^2} |10\rangle$ for some $\alpha \approx 0.5$, thus increasing population swap contrast. The bond kinetic energy measurement is calibrated to have maximum contrast, so we can directly fit a sinusoid to the population swaps of the resonantly ramped state $|10\rangle + |01\rangle$.

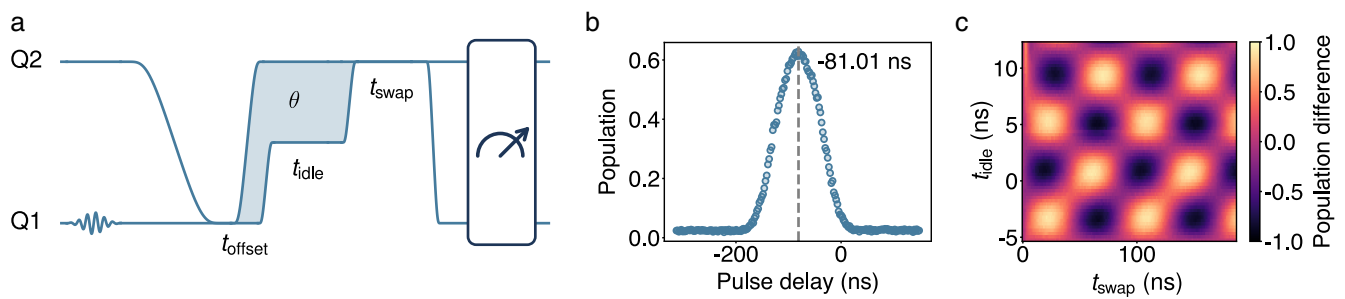

FIG. S5. Beamsplitter timing calibration. **(a)** The experimental sequence to fine-tune the acquired phase $\theta$ between the two qubits involved in the beamsplitter interaction. With $t_{\mathrm{idle}}$ and the frequency difference as control knobs, $\theta$ can be arbitrarily tuned, thus implementing any linear combination of effective $\sigma_x$ and $\sigma_y$ measurements. **(b)** Calibration of the delay of the arrival of a fast-flux pulse relative to a qubit drive pulse on a qubit. The $x$-axis indicates the scheduled time delay between the fast-flux pulse and the qubit drive. The peak of the curve indicates the optimal delay to align the two pulses in time. **(c)** A plot of the population difference between the two qubits showing resonant qubit swaps with $t_{\mathrm{swap}}$ swept on the $x$-axis and $t_{\mathrm{idle}}$ swept on the $y$-axis. Tuning $t_{\mathrm{idle}}$ changes the phase accumulated between the qubits, thus changing the contrast of resonant swaps.

## S7. BEAMSPLITTER MEASUREMENT

Following evolution under a double-well Hamiltonian $H = \hbar J(a_l^\dagger a_r + a_r^\dagger a_l)$, the particle number difference of two coupled resonant harmonic oscillators can be expressed as [9, 10]

$$n_r(t) - n_l(t) = \cos(2Jt)\left[n_r(0) - n_l(0)\right] + \mathrm{i}\sin(2Jt)\left[a_l^\dagger(0)a_r(0) - a_r^\dagger(0)a_l(0)\right]. \tag{S4}$$

Given an evolution time $t_{\mathrm{BS}} = \pi/4J$, a measurement of the occupation difference translates to a measurement of the current operator on the initial state:

$$n_r(t_{\mathrm{BS}}) - n_l(t_{\mathrm{BS}}) = \mathcal{J}_{l\rightarrow r}(0)/J. \tag{S5}$$

This measurement protocol detects currents in the two-qubit subspace $\{|00\rangle, |01\rangle, |10\rangle, |11\rangle\}$, with $|01\rangle$ and $|10\rangle$ corresponding to currents of $+J$ and $-J$, and with $|00\rangle$ and $|11\rangle$ both corresponding to zero current. While, in principle, there exist additional currents corresponding to higher bosonic modes of the transmon, with our device parameters for $U$, $J_\parallel$, and $J$, we expect a significant portion of the current to be restricted to the two-qubit computational basis above. Figure S7 highlights the discrepancy between measuring the current operator directly or through the mapping to particle number imbalance.

## S8. TLS SPECTROSCOPY

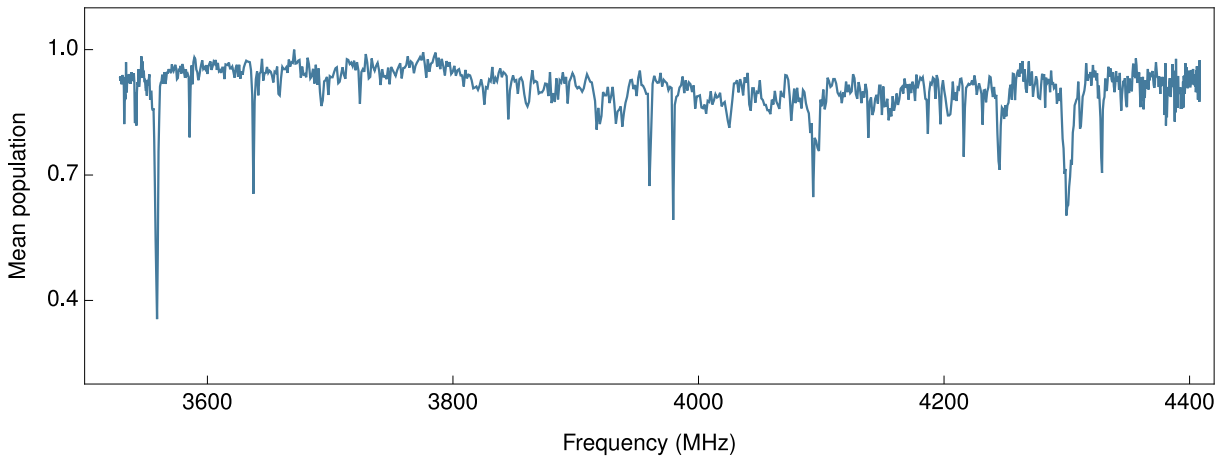

FIG. S6. TLS spectroscopy of the qubit mode. Excited-state population of the qubit measured after a fixed delay as a function of qubit frequency. Narrow dips in the population indicate enhanced relaxation at frequencies where the qubit hybridizes with strongly coupled environmental two-level systems (TLSs).

As part of the process for calibrating the adiabatic ramp, we find the qubit $T_1$ lifetimes depend sensitively on frequency-dependent environmental effects. We attribute lower lifetimes to the presence of lossy two-level systems (TLSs) coupled to the qubit mode [11, 12]. Strongly-coupled TLSs appear as sharp dips in $T_1$ at qubit frequencies near the TLS frequency. We characterize the frequency landscape by measuring qubit populations after a short delay at various frequency points. Frequencies for which the qubit interacts with a TLS exhibit faster population decay than is expected from the baseline $T_1$ lifetime. For consistency, we employ fast-flux pulses to tune the qubit initialization and readout frequencies to be the same throughout the experiment.

A trace of the qubit population versus frequency is shown in Fig. S6. For each qubit, we choose the frequencies for driving, adiabatic ramps, beamsplitter operations, and readout that minimize population decay from TLSs. Furthermore, we design the adiabatic ramp to tune the frequency of only the sites without excitations, in order to reduce the range in frequency space in which the qubit might interact with lossy environmental modes.

## S9. STATE PREPARATION AND ADIABATIC RAMP

Our state-preparation scheme involves tuning the frequencies of all qubits onto resonance ($\omega_{\mathrm{lattice}}$). To prepare the highest half-filling eigenstate, we excite the four highest-frequency qubits to $|1\rangle$ while they are detuned by more than

$J$ and $J_\parallel$. We pick qubits 1, 4, 5, and 8 as the excited qubits to minimize effects of $ZZ$ crosstalk and level crossings with $|2\rangle$ states. We diabatically tune the four excited qubits to $\omega_{\text{lattice}}$ and simultaneously tune the four unoccupied sites to $\omega_{\text{lattice}} - 25J$. This state remains the highest-energy eigenstate, since the occupied and unoccupied sites have a separation much larger than $J$ and $J_\parallel$. Slowly ramping the resonance frequency of the four unoccupied sites adiabatically prepares the highest excited state of the Hamiltonian with negative $U$, and thus effectively prepares the half-filling ground state of the Hamiltonian with positive $U$ and $\pi$-shifted plaquette fluxes as seen in Eq. (2) of the main text. This state-preparation procedure avoids tuning the frequencies of the excited qubits to suppress particle loss through level crossings with environmental decay channels (see Sec. S8 for more details).

We find that the energy levels, throughout the adiabatic ramp, depend sensitively on which qubits are excited. We confirm through exact diagonalization and simulations of adiabatic ramps that the state with particles on sites 1, 4, 5, and 8 remains sufficiently gapped throughout the adiabatic evolution. We choose a ramp time of 290 ns, balancing the requirement for adiabatic evolution set by the energy-level gaps against the finite $T_2$ coherence times.

## S10. READOUT

We determine the state of each qubit using the qubit-state-dependent dispersive shift of its readout resonator's frequency [2], performing heterodyne detection to measure the amplitude and phase of our probe signal as it passes through the transmission line on our device. We calibrate single-shot readout before each experimental run by performing 8000 experiments measuring the averaged in-phase ($I$) and quadrature ($Q$) signals when the qubit is in the ground state and another 8000 when the qubit is in the first excited state. We then pick a discriminator line in the $I$-$Q$ plane that maximizes the correct assignment of prepared ground- and excited-state points to the measured ground ($|0\rangle$) and excited ($|1\rangle$) states.

Single-shot readout fidelities for all qubits typically exceed 90% if the qubit frequencies are at maximal detuning from coupler frequencies ($\Delta_{qc} \approx 3$ GHz). On the other hand, we notice that as detunings decrease, single-shot discrimination fidelity worsens, likely due to increased qubit-coupler hybridization and worse coupler $T_1$ times. For coupler-qubit configurations used in our experiments, single-shot readout fidelities are generally between 70% and 90%.

Current correlation measurements require simultaneous readout of four qubits, which is done by multiplexing readout tones at the frequencies of each corresponding resonator. To improve the fidelity of our data, we postselect by retaining only experiments in which exactly four excitations are measured across the lattice, correcting out decay errors. Therefore, we read out all eight qubits on our lattice simultaneously.

## S11. NUMERICAL SIMULATIONS OF THE TRIANGULAR LADDER

We perform simulations of state preparation and beamsplitter operation by modeling our system according to the Hamiltonian in Eq. (1) of the main text. All simulations in this work use the QuTiP package [13]. We restrict the Hilbert space to include only states with exactly four particles on the eight sites and obtain eigenstates with exact diagonalization. We use measured values for the coupling strengths $J_\parallel$ and $J$ and for the interaction strength $U < 0$. Following the transformation in Eq. (2) of the main text, we obtain the eigenstate with the highest energy, which corresponds to the ground state of the Hamiltonian with positive $U$ and $\pi$-shifted plaquette fluxes. Since the adiabatic ramp duration is itself a non-negligible fraction of the lowest $T_2$ lifetimes, we also simulate an adiabatic ramp with dephasing to compare the prepared state to the exact eigenstate. We simulate the dynamics during the ramp using the Lindblad master equation

$$\dot{\rho} = i[H, \rho] + \sum_j^N \gamma_j \left( L_j \rho L_j^\dagger + \frac{1}{2} \{L_j L_j^\dagger, \rho\} \right) \tag{S6}$$

with

$$L_j = \sqrt{2} a_j^\dagger a_j \tag{S7}$$

as the collapse operator corresponding to qubit dephasing and $\gamma_j$ the dephasing rate.

After thus obtaining the state prepared during the ramp, we extract both the expectation value of the current as well as the population difference during the beamsplitter interaction. We incorporate the exact detunings and onsite interaction strengths of qubits to match our experimental procedure as best as possible, finding that these factors reduce the measured current correlations from their ideal values.

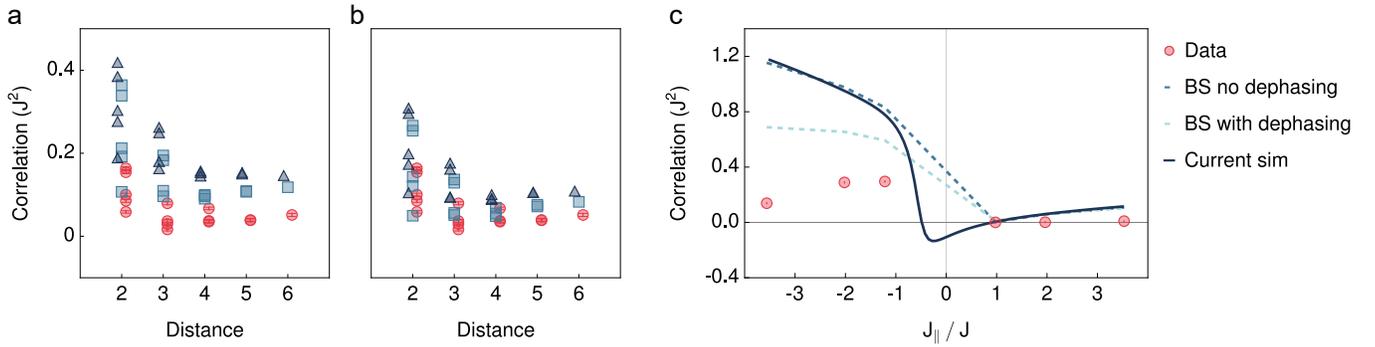

FIG. S7. **(a)** Comparison of experimentally measured current-current correlations as a function of rung separation with numerical simulations. Red circles denote experimental data. Dark blue triangles show $\langle \mathcal{J}_i \mathcal{J}_j \rangle$ in the ground state computed using exact diagonalization. Light blue squares represent the expected population-difference correlations obtained by simulating the beamsplitter measurement applied to the exact ground state. Points are offset along the $x$-axis for visibility. **(b)** Same comparison as in (a), but incorporating dephasing during state preparation. Current operator and beamsplitter measurement expectation values are calculated with the simulated mixed state after the adiabatic ramp. **(c)** Chiral-current order parameter $\mathcal{C}$ as a function of the rung-to-leg coupling ratio $J_\parallel / J$. Direct operator simulation of the exact ground state, as well as simulation of the beamsplitter measurement with and without dephasing are shown for comparison.

While simulating unitary time evolution on the state in the four-particle subspace is straightforward, the post-ramp state is generally a mixed state which requires Lindbladian dynamics. To greatly simplify the simulations, we can recognize that pairs of qubits are well separated in frequency during the beamsplitter interaction, so we can use the reduced density matrix of each pair in simulating the population difference measurement. Assuming there are no more than four particles in total, we truncate the two-qubit bosonic space to states with at most four particles per site.

Figures S7 and S8 compare the measured current–current correlations and bond kinetic energies with numerical simulations. In each case, we show both the correlations obtained from direct operator expectation values and those inferred from a simulation of the beamsplitter measurement protocol. We find that the agreement between experiment and simulation improves when dephasing during state preparation is included in the model. Nevertheless, the simulations still overestimate the measured signal amplitudes, indicating that the model does not capture all relevant decoherence mechanisms. Incorporating additional dissipative processes, beyond simple dephasing, may therefore provide a more accurate description of the experimental dynamics.

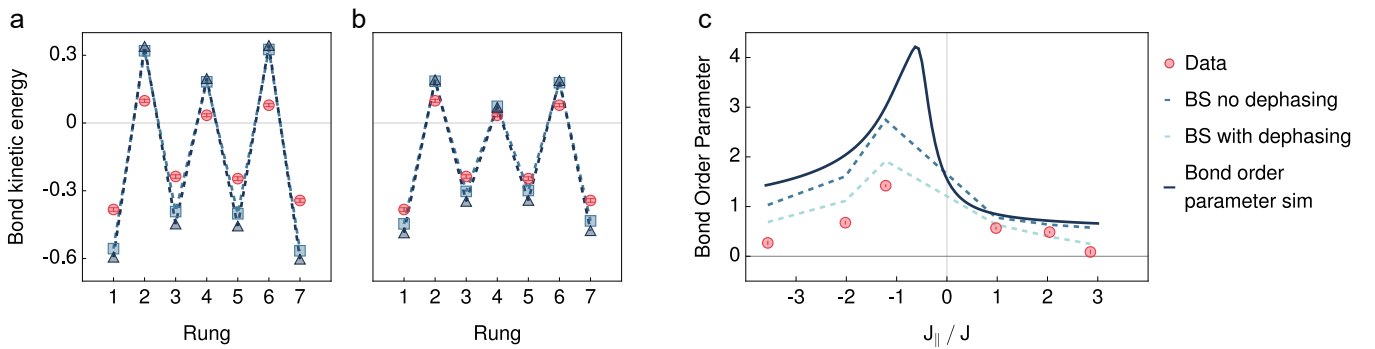

FIG. S8. **(a)** Comparison of experimentally measured bond kinetic energies with numerical simulations. Red circles denote experimental data. Dark blue triangles show $\langle \mathcal{O}_j \rangle$ in the ground state computed using exact diagonalization. Light blue squares represent the expected population difference obtained by simulating the beamsplitter measurement applied to the exact ground state. Points are offset along the $x$-axis for visibility. **(b)** Same comparison as in (a), now including dephasing during state preparation. Red circles again represent the experimental data. Dark blue triangles denote bond kinetic energies extracted directly from the simulated mixed state after state preparation, while light blue squares show those obtained after simulating the beamsplitter measurement. **(c)** The bond order parameter $O_{BO}$ as a function of the rung-to-leg coupling ratio $J_\parallel / J$. Direct operator simulation of the exact ground state, as well as simulation of the beamsplitter measurement with and without dephasing are shown for comparison.

ADDITIONAL DATA ON CURRENT CORRELATIONS

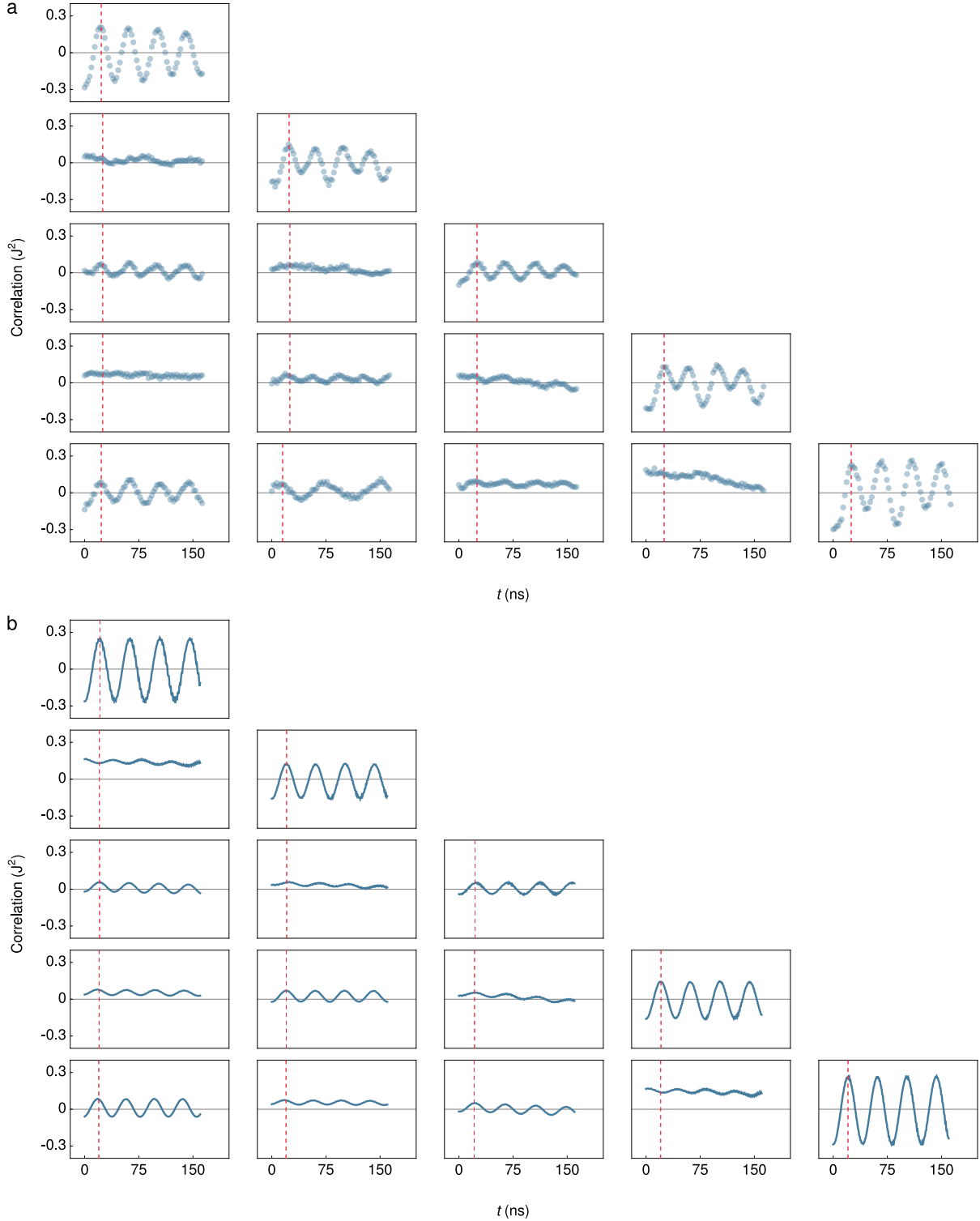

FIG. S9. Beamsplitter measurement for current-current correlations at $J_\parallel/J = -2$. **(a)** Traces of measured rung-rung number difference correlations over time, during evolution under $H_{\mathrm{BS}}$, for all pairs of rungs. Axes represent the same rungs as in the correlation matrices in Fig. 3b,c of the main text. The red dashed line indicates $t_{\mathrm{BS}}$, when the population-difference correlations correspond to current-current correlations. **(b)** Simulated beamsplitter measurements of current-current correlations.

## ADDITIONAL DATA ON THE BOND ORDER PARAMETER

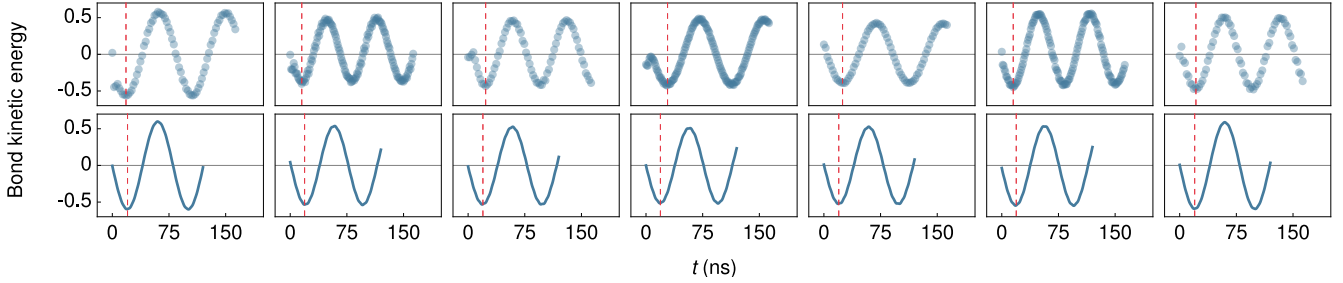

FIG. S10. Beamsplitter measurement for bond kinetic energy at $J_\parallel/J = 1$. Time traces of particle-number differences are plotted against evolution time under $H_{\mathrm{BS}}$. Plots correspond to the rungs in Fig. 4d,e of the main text. The two qubits of each rung acquire a relative phase of $\pi/2$ to enable the kinetic energy measurement. The red dashed line indicates $t_{\mathrm{BS}}$, when the population difference corresponds to the bond kinetic energy.



\end{document}